\documentclass[aps,pra,twocolumn,showpacs,superscriptaddress,preprintnumbers,amsmath,amssymb,floatfix,fnpara,nofootinbib]{revtex4}
\usepackage{graphicx}                                                                           
\usepackage{dcolumn}                                                            
\usepackage{bm}                                                                                            
\usepackage{color}                                                                                      
\def\etal{{\it et al }}
\begin{document}
\author{N.~R. Badnell\email[]{badnell@phys.strath.ac.uk}}
\affiliation{Department of Physics, University of Strathclyde, Glasgow G4 0NG, United Kingdom}
\author{C.~P. Ballance}
\affiliation{Department of Physics, Auburn University, Auburn, Alabama 36849}
\author{D.~C. Griffin}
\affiliation{Department of Physics, Rollins College, Winter Park, Florida 32789}
\author{M. O'Mullane$^1$}
\title{Dielectronic recombination of W$^{20+}\, (4{\bf d}^{10}4{\bf f}^{8}$): addressing the half-open f-shell}
\date{\today}
\begin{abstract}
A recent measurement of the dielectronic recombination (DR) of W$^{20+}$ 
[Schippers \etal Phys. Rev. A {\bf 83}, 012711 (2011)]
found an exceptionally large contribution from near threshold resonances ($\lesssim 1$~eV). 
This still affected the Maxwellian rate coefficient at much higher temperatures.
The experimental result was found to be a factor 4 or more than that currently
in use in the $100-300$eV range which is of relevance for modeling magnetic fusion plasmas. 
We have carried-out DR calculations with {\sc autostructure} which include all
significant single electron promotions. Our intermediate coupling results are more than a factor of 4
larger than our $LS$-coupling ones at 1~eV but still lie a factor 3 below experiment here.
If we assume complete (chaotic) mixing of near-threshold autoionizing states then
our results come into agreement (to within 20\%) with experiment below $\lesssim 2$~eV.
Our total IC Maxwellian rate coefficients are 50--30\% smaller than those based-on experiment over 100--300~eV.

\end{abstract}
\maketitle

\section{Introduction}
\label{In}

Tungsten will be a key element~\cite{Putt} in the ITER magnetic fusion device~\cite{ITER1,ITER2,ITER3} currently 
under construction at Cadarache in France~\cite{Cad}. Its ability to withstand high power-loads means that it will be
the primary facing material within the vacuum vessel. Its high nuclear charge means also that it is
potentially a serious contaminant in the sense of its ability to quench the fusion reaction
due to radiative power loss. Intensive studies are underway at all of the world's major magnetic
fusion laboratories to understand, predict, and control its behavior. The recent ITER-like wall
upgrade at JET~\cite{JET} provides the closest reactor environment short of ITER itself.
The 74 ionization stages of tungsten may seem daunting from the detailed theoretical perspective.
Reality is somewhat different. Very few ionization stages are observed in practice.
Near neutrals are seen as they sputter-off surfaces. Then many ionization stages burn through
quite rapidly before ions of much higher charge-state are observed in particular localized
environments. W$^{20+}$ is a  significant stage for spectral diagnostics and is seen at the null point of the
separatrix at JET. W$^{44+}$ performs a similar function at the core and is observed by the JET KX1 spectrometer. 
The ionization stages will change with the much larger and hotter ITER device but the principle remains the same:
very few stages need to be modeled in detail. The great bulk of them can be modeled more `coarsely' as superstages.
Detailed studies are being made of these key stages. One of the most basic and important theoretical quantities
is the tungsten ionization balance since it is the main determinant of the intensity of spectral line emission. 
Electron-collisional equilibrium is a balance between
electron-impact ionization and dielectronic recombination. (All other recombination processes are
negligible in the magnetic fusion domain.) A sufficiently accurate theoretical description
of dielectronic recombination is key. 

A recent experiment on  W$^{20+}$ ions by Schippers \etal~\cite{Sch} at the TSR storage ring at 
Heidelberg measured an exceptionally large dielectronic recombination merged-beams `rate coefficient' at a few eV.
So-much-so that its contribution at the temperatures of significant fractional abundance for W$^{20+}$ (100--300~eV)
gave rise to a Maxwellian rate coefficient that is a factor 4 or more larger than that currently
used by the main magnetic fusion modeling package --- the Atomic Data and Analysis Structure (ADAS)~\cite{ADAS}.
We seek to resolve this discrepancy.

There is little previous detailed work on DR for f-shell ions. 
At one end ($4d^{10} 4f$) there are calculations for Gd$^{17+}$ by Dong \etal~\cite{Dong} utilizing the
Flexible Atomic Code (FAC)~\cite{FAC} which are relevant for modeling soft X-ray lithography. At the other end 
($4d^{10} 4f^{13+}$) there are calculations for Au$^{20+}$ by Ballance \etal~\cite{Ball} with {\sc autostructure}~\cite{AS}. 
The results of Ballance \etal are in good agreement with the measurements of Schippers \etal~\cite{Sch_Au}
from 2~meV up to 10~eV. Both of the above approaches are standard level-resolved calculations
which allow for single electron promotions-plus-capture. They are largely restricted to at most
doubly-excited configurations and interactions thereof. 

Previous work on open f-shell ions is apparently limited to the configuration average approximation,
the Burgess General Program (BBGP)~\cite{Burg, BBGP}, and others of that ilk which are the mainstay of modeling codes.
The calculations of Flambaum \etal~\cite{Flam1} for the DR of Au$^{25+}$ (which is isoelectronic
with W$^{20+}$) can be viewed as a form of partitioned configuration average. They utilize
expressions for the radiative rate and autoionization rate which are similar to those of
the configuration average. The near-threshold autoionization rates are partitioned
over a Breit-Wigner distribution which is characterized by a single spreading width~\cite{Gleb1}.
This compares with our previous configuration average work~\cite{Bad89} which partitions them
over the non-metastable core levels according to their statistical weight.
Such a partitioning maintains the allowed- vs forbidden-channel nature which is characteristic 
of DR in simple systems.
The justification of the Breit-Wigner form and the spreading width follows from the complexity of the open f-shell 
ions in which configuration mixing tends to a chaotic limit which can be described by statistical theory~\cite{Flam2}.
This leads to a structureless continuum for the near-threshold merged-beams DR rate coefficient.
The Au$^{25+}$ results of Flambaum \etal~\cite{Flam1} were found to be in good agreement with the measurements
of Hoffknecht \etal~\cite{Hoff} below 1~eV.

The approach to DR reported-on in this paper is a detailed (level-to-level) one with single
electron promotions-plus-capture for all significant contributing configurations. The configuration mixing
(and spin-orbit mixing) that we allow-for is the same as done in our previous work on the open Fe M-shell 
($3p^q$)~\cite{iron} and the open Sn N-shell ($4d^q$)~\cite{tin} ions. This allows for the mixing of all 
autoionizing configurations within the $(N+1)$-electron complex when the promoted and captured electrons 
have the same  principal quantum number. The inequivalent case restricts the mixing to configurations
of the $N$-electron core i.e. with common Rydberg $nl$ quantum numbers. This approach gives
rise to a near-structureless continuum for the near-threshold merged-beams DR rate coefficient
for these Fe and Sn cases. We seek to extend this work to the half-open f-shell and to compare the 
near-threshold merged-beams DR rate coefficient with the measurements of Schippers \etal~\cite{Sch}. 
We seek also to determine the high energy Maxwellian DR rate coefficient applicable to the diagnostic 
modeling of W$^{20+}$ in magnetic fusion plasmas.

We provide a description of our background theory in Sec.~\ref{The}. 
We describe its application to W$^{20+}$ in Sec.~\ref{App}. 
We present out results and compare them with those of experiment in Sec.~\ref{Res}.
We make some concluding remarks in Sec.~\ref{Con}.

\section{Theory}
\label{The}

We use the independent processes and isolated resonance approximations to describe dielectronic
recombination~\cite{Burg}. The oft-repeated working equations may be found in~\cite{iron}
along with a more expansive discourse on the methodology we employ. We summarize some pertinent points.

\subsection{Methodology}
\label{Meth}

We use the computer code
{\sc autostructure}~\cite{AS} to calculate all relevant atomic parameters: energy levels,
radiative rates, and autoionization rates. A multi-configuration expansion is used in an
$LS$-coupling or intermediate coupling (IC) representation. The configuration average (CA) 
representation~\cite{CA1, CA2} is a simpler approach which is very useful for complex heavy species 
since it provides a rapid overview of the problem. 

All of our previous works with the CA
approximation utilized the {\sc dracula} code~\cite{CA2,CA3}. One immediately sees the effect of level-resolution
and fine-structure mixing when comparing $LS$-coupling with IC results obtained with {\sc autostructure}. 
Comparison of $LS$-couping with CA results has been clouded by the fact that
{\sc dracula} is based upon the Cowan structure code~\cite{Cow}. The Cowan code utilizes kappa-averaged
Hartree-Fock radial orbitals~\cite{CG1}. The differences here with {\sc autostructure} can be minimized by its 
use of kappa-averaged orbitals computed in self-consistent configuration average model potentials~\cite{Cow}.
The Cowan code generally also scales various operator interactions\footnote{It does not do so under CA operation.}. 
This facility is not readily or generally
available in {\sc autostructure}. We have implemented the CA angular momentum representation
within {\sc autostructure}. This eliminates any uncertainty in seeing the pure effect of moving to
a configuration-mixed term-resolved representation. We have carried-out detailed CA comparisons
between {\sc autostructure} and {\sc dracula} in the course of the present work so as to verify
the integrity of the new development.

\subsection{Computation}
\label{Comp}

The (near-) half-open f-shell problem is a daunting one. If we view it simply in terms of binomial
coefficients for the number of states present in a configuration then moving from the $4d^{10}4f^{13}$
ground configuration considered in~\cite{Ball} to $4d^{10}4f^8$ increases the number of states
by a factor (429/2). Memory requirements (CPU and disk) scale as $(429/2)^2$ and the time requirement
as $(429/2)^3$. This is all relative. Absolute numbers are much larger once we start promoting electrons
from the $4d-$ and $4f-$ subshells of the ground configuration.
(The only bonus of the binomial effect is that $4f^7$ is only marginally worse than $4f^8$ since the
number of states increases only by a factor 8/7.)

The published {\sc autostructure} code~\cite{AS} was used for our recent work on the tin half-open 
d-shell~\cite{tin} and Au$^{20+} 4f^{13}$~\cite{Ball}. It does not scale to the half-open
f-shell. Substantial development has been necessary. A detailed exposition is more suited to
a computer physics journal and so we give only a flavor here. 

Most angular momentum packages
used by atomic structure and collision codes are based upon Racah algebra. This is a hierarchical
coupling scheme. A complication for the open f-shell is the need to introduce a new quantum
number: seniority. {\sc autostructure} employs the non-hierarchical Slater-state approach to
angular momentum coupling that was advocated by Condon and Shortley~\cite{TAS}. It has
no concept of parentage\footnote{It is both possible and advantageous to introduce some
parentage but it is not required.}. All interactions are expanded and determined initially
in an uncoupled (Slater-state) representation. A transformation is then made to an $LS$ and/or $LSJ$
representation through the use of vector-coupling coefficients (VCCs). The half-open f-shell requires
billions of coefficients and so tens of Gbytes of RAM per processor. They are not all required at
the same time. It has been possible to implement an archival on disk in such a way that does
not swamp the calculation with I/O. 

Another issue concerns the large number of radiative and
autoionization rates that arise in a level-resolved calculation. The approach to-date has been
to archive them all to disk for subsequent processing in a variety of ways: to compare
with experiment, to generate total rate coefficients for astrophysical modeling, or to 
process as final-state resolved data for collisional-radiative modeling for magnetic fusion.
Such general flexibility comes at the cost of many Tbytes of disk space and corresponding I/O time.
If we sacrifice some degree of generality then we can carry-out bundling over quantum numbers and
summation of partial widths on-the-fly as the atomic data is generated. This reduces the
data files to a manageable size. The user choice of bundling and/or summation
should be guided by the exact same implementation made within the collisional-radiative modeling
approach so as to render it tractable for heavy species. It is important to note that this
introduces {\it no} additional approximation for our description of the experimental DR cross
section nor the total Maxwellian rate coefficients presented later.

We do not dwell on various RAM and CPU time issues that arose as well on scaling from the
open d- to f-shell. It is sufficient to note that  the calculations
reported-on below took about two weeks on a modest cluster (the problem does not engender large-scale 
parallelization) with 4Gb RAM and 250Gb of scratch disk per processor.

\subsection{Mixing}
\label{Mix}

We  discuss the role of (configuration) mixing of autoionizing states on DR in complex heavy species.
This is important since we include only a limited amount (see the next section).
Consider a model problem in which we include only autoionizing configurations which result
from single electron promotions-plus-capture from the the ground configuration. We assume
that the autoionization rates ($A_1$) and radiative rates ($R$) satisfy $A_1>>R$. Then the DR cross section 
is basically proportional to $R$ for a fixed symmetry. (We consider only autoionizations which are inverses 
of the dielectronic capture in this model i.e. near-threshold.) Now consider the addition of autoionizing
configurations which result from double electron promotions-plus-capture from the ground configuration.
This set is only populated by mixing with the first via a unitary transformation. 
We denote the second set of autoionization rates by $A_2$ and
assume similar radiative rates ($R$). We assume $A_2<<R$. If mixing between the two sets is strong enough and there
are enough states then the autoionization rates for the first set (which we label now as $A_{12}$) are depleted to the
extent that $A_{12}<<R$. The DR cross section is now proportional to $A_{12} + A_2 = A_1$.  The total is
{\it enhanced} by a factor $A_1/R$. (Mixing only
takes place between states of the same statistical weight and so the sum over autoionizing states is 
assumed to be implicit.) We consider the effect of further mixing. 
Now add configurations which result from triple electron promotions-plus-capture from the ground configuration.
Denote the autoionization rates for the three sets as $A_{123}$, $A_{23},$ and $A_{3}$. Assume all $A_i<<R$ still.
The total DR cross section is unchanged since it is proportional to $A_{123}+ A_{23} + A_3 = A_1$ still.
Such mixing is merely {\it redistributive} and so can be neglected with respect to total DR.

This is the nature of the near-threshold DR problem in complex ions described by Flambaum \etal~\cite{Flam1}
(for Au$^{25+}$) following a preliminary earlier study~\cite{Gleb1}. The chaotic fully-mixed nature of Au$^{24+}$ was 
verified by Gribakin and Sahoo~\cite{Gleb2} who performed large-scale calculations of eigen-energies and eigen-vectors.
Our detailed description of the autoionizing spectrum is necessarily incomplete.
The above discussion is intended to shed light on how far we need to go. 
The model problem was merely illustrative in using configuration mixing via different sets of promotions.
We are largely restricted to one electron promotions-plus-capture since we need to compute autoionization
rates and radiative rates which are applicable over a much wider range of energies.
But we do have differing representations: CA, $LS$-coupling and IC. The question then is one of the
degree to which the IC representation is incomplete with regards to enhancement or whether it has moved
to a redistributive regime so that our DR cross sections will have converged largely with
respect to the total. The structured behavior of our low energy cross section, its comparison with
experiment, and with statistical theory will enable us to judge the degree to which this is so.
(It is straightforward to apply the statistical  approach to our data --- we simply partition our autoionization
rates using the Breit-Wigner distribution --- but note that this procedure is only valid near threshold.)

\section{Application}
\label{App}

The ground configuration of W$^{20+}$ is [Kr]$4d^{10}4f^{8}$. The ground term is $^7F$. The ground level is $J=6$.
These denote the ground state of the CA, $LS$-coupling, and IC representations.
We describe the DR reactions that we take account-of by their configuration representation.
This consists of $N$-electron configurations to which a continuum electron and a Rydberg electron
are each coupled. This describes dielectronic capture and/or autoionization. The latter describes
radiative transitions within the core also. Additional $(N+1)$-electron `correlation' configurations are added
to describe Rydberg electron radiation into the core. Rydberg--Rydberg radiation $n\rightarrow n'>4$
is described hydrogenically. 
We breakdown the problem by target subshell promotion.

\subsection{$\Delta n=0$}
We allow for $4d\rightarrow 4f$ and $4f\rightarrow 4f$ promotions in both $LS$-coupling and IC.
The latter promotion does not contribute to CA DR since it corresponds to an elastic transition.
The $N$-electron configurations are $4d^{10}4f^{8}$ and $4d^{9}4f^{9}$. 
The $(N+1)$-electron configurations are $4d^{10}4f^{9}$ and $4d^{9}4f^{10}$.
We note that some terms/levels of $4d^{10}4f^{7}5l$ lie below $4d^{9}4f^{9}$.
The former could provide an alternative autoionization channel for the latter.
Such a transition is forbidden directly since it is described by a two-body operator.
(A continuum electron is still to be coupled to the former and a Rydberg to the latter.)
It could take place via mixing if we were to include additional configurations.
We do not.
We consider also the $4p\rightarrow 4f$ promotion in CA only. Its contribution is
expected to be small.
We normally use the CA calculation to determine the range of Rydberg-$nl$ required
to converge the total DR and then use these values subsequently for the more demanding
$LS$-coupling and IC calculations. We find that the contribution to the Maxwellian rate coefficient 
from the $4d\rightarrow 4f$ promotion is converged to about 3\% by $n=100$ and $l=7$ at all energies.
It is not possible to do so for the $4f\rightarrow 4f$ promotion
since the CA result is zero and so we utilize the $LS$-coupling calculation here to delimit the IC one.
We find that the contribution from the $4f\rightarrow 4f$ promotion is converged to about 3\% by 
$n=100$ and $l=6$ at all energies.

\subsection{$\Delta n=1$}
We consider the $4d$ and $4f$ promotions separately.

\subsubsection{$4f\rightarrow 5l$}
The $N$-electron configurations are $4f^8$ and $4f^7 5l\,(l=0-4)$.
It is necessary to omit the $n=5$ continuum so as to keep the
problem tractable in the $LS$-coupling and IC  
calculations\footnote{The number of VCCs that need to be internally buffered becomes too large. A smaller buffer
could be implemented but this would likely increase I/O time substantially. The absolute number of VCCs
required in $LS$-coupling is only typically a factor of 2 smaller than for IC. The demands
of the IC calculation arise from the fact that far more of the states that they represent interact.}.
We carried-out CA calculations both with and without the $n=5$ continuum to aid our analysis
of the uncertainty (overestimate) in our $LS$-coupling and IC results. There is none at all
below $\sim 20$~eV since they are all energetically closed.

The $(N+1)$-electron configurations are $4f^8 5l\,(l=0-4)$.
Some of the $4f^7 5l5l'$ configurations are (partially) bound. We treat such $n\rightarrow n'=5$
radiation hydrogenically (approximately) in the $LS$-coupling and IC calculations.
They are either strictly bound or autoionizing in the CA approximation.
Their contribution is small.  

The CA results for this promotion are converged to about 5\% at 200eV and 10\% at 1000~eV on summing to $l=5$.
The sum over $n$ is converged to better than 2\% by $n=100$.
We add this small `top-up' in $l$ (and $n$ since we are doing so) from our CA results to the $LS$-coupling and IC ones.

\begin{figure}[tbp]
\includegraphics[scale=0.8, angle=-90]{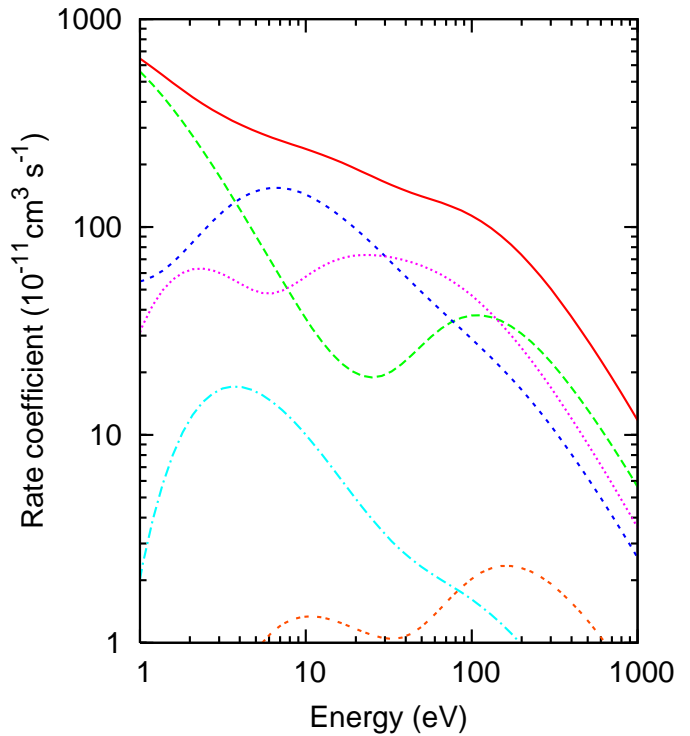}\vspace*{-4mm}
\caption{\label{Fig1}
(Color online) W$^{20+}$ CA Maxwellian DR rate coefficient contributions for various promotions:
total (solid red curve), $4d\rightarrow 4f$ (long-dashed green curve), $4f\rightarrow 5l$ (short-dashed blue curve),
$4d\rightarrow 5l$ (dotted magenta curve), $4p\rightarrow 4f$ (dot-dashed cyan curve), and
$4d+4f\rightarrow 6l$ (double-dashed orange curve).}
\end{figure}

\subsubsection{$4d\rightarrow 5l$}
The $N$-electron configurations are $4d^{10}4f^8$, $4d^{9}4f^9$, $4d^{9}4f^8 5l\,(l=0-4)$ and $4d^{10}4f^7 5l\,(l=0-4)$.
We omit the $n=5$ continuum again. The $4d^{9}4f^8$ is rather demanding when coupled to $5lnl'$ for $l+l'>5$.
It has a factor 70/8 more states than the corresponding $4d^{10}4f^7$. We need to consider it further.
We write the dielectronic capture reaction in a somewhat unusual form:
\begin{eqnarray}
4f^8(^{7}F)4d^{10} + {\rm{e}}^{-} & \rightarrow & 4f^8(^{7}F)4d^{9}5lnl'\,.\nonumber
\end{eqnarray}
This illustrates the role of the $4f^{8}\,^{7}F$ ground term as a spectator. It cannot change
simultaneously with the two-body dielectronic capture. It can change (shake-up) via mixing in the 
autoionizing states. We omit such mixing. We do the same for the reverse radiative stabilization to
$4f^8(^{7}F)4d^{10}nl$. This renders a tractable but reasonable description of the $4d\rightarrow 5l$ promotions. 
We recall that there is no configuration mixing whatsoever present in the CA approximation.
We recall also the lack of sensitivity to configuration mixing that we found for total DR in the tin $4p-4d$
transition array despite the demonstrable extensive configuration mixing~\cite{tin}.
Such an argument may not be valid near threshold --- see the discussion in Sec.~\ref{Mix}.
We emphasize that we place no such restrictions (on mixing) when the $4f$ is active such as for the $4d^{9}4f^9$ configuration.
We have implemented the general user specification of term restrictions for spectator subshells within {\sc autostructure}.
These are common to all configurations which contain the specified subshell(s).

The $(N+1)$-electron configurations are $4d^{10}4f^8 5l\,(l=0-4)$ and $4d^{9}4f^9 5l\,(l=0-4)$.
A few of the $4d^{9}4f^8 5l5l'$ configurations are (partially) bound and we treat them as for {$4f\rightarrow 5l$}.

The CA results for this promotion are converged to 1\% at 1000~eV on summing to $l=4$.
The sum over $n$ is converged to much better than 1\% by $n=100$.

\begin{figure}[tbp]
\includegraphics[scale=0.8, angle=-90]{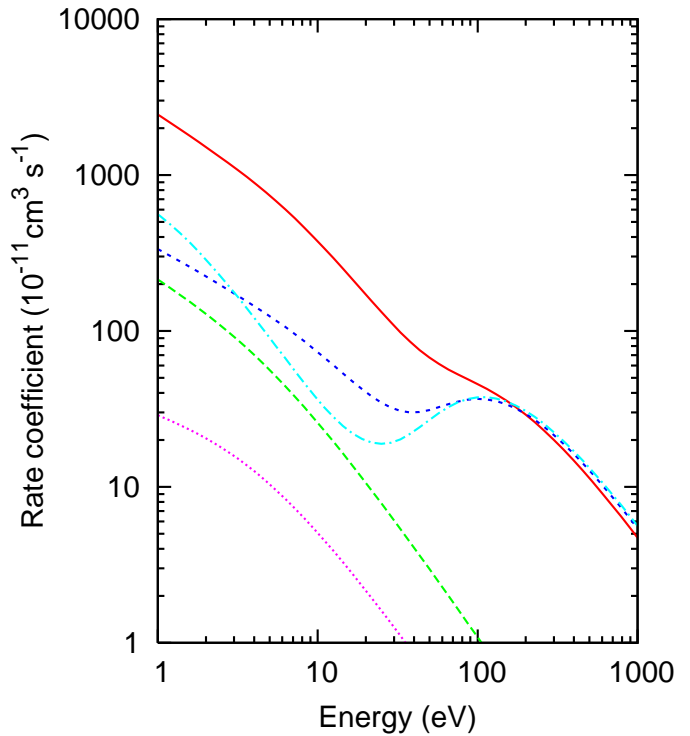}\vspace*{-4mm}
\caption{\label{Fig2}
(Color online) W$^{20+}$ Maxwellian DR rate coefficient contributions for $\Delta n=0$ promotions:
IC $4d\rightarrow 4f$ (solid red curve), IC $4f\rightarrow 4f$ (long-dashed green curve), LS $4d\rightarrow 4f$
(short-dashed blue curve), LS $4f\rightarrow 4f$ (dotted magenta curve), and CA $4d\rightarrow 4f$ (dot-dashed cyan curve).}
\end{figure}

\subsection{$\Delta n=2$}
We consider $4d+4f\rightarrow 6l$ promotions within the CA approximation only.
The contribution is expected to be small.

\section{Results}
\label{Res}

We show an overview of the different CA contributions to the total DR Maxwellian rate coefficient in Fig.~\ref{Fig1}.
The energy range of interest for an electron collisional plasma is 100--300~eV.
This is where W$^{20+}$ has its maximal fractional abundance ($>0.01$) in a magnetic fusion plasma.

We remark that the contribution from $4d\rightarrow 4f$ promotions is comparable with the $\Delta n=1$
above $\sim 100$~eV. This is in contrast to case of the almost full 4f-subshell case of Au$^{20+}$~\cite{Ball}.
It also dominates at a few eV but this behavior can be expected to be more ion-dependent.

We see that we do not need to consider $4p\rightarrow 4f$ and $\Delta n=2$ promotions any further.

\begin{figure}[tbp]
\includegraphics[scale=0.8, angle=-90]{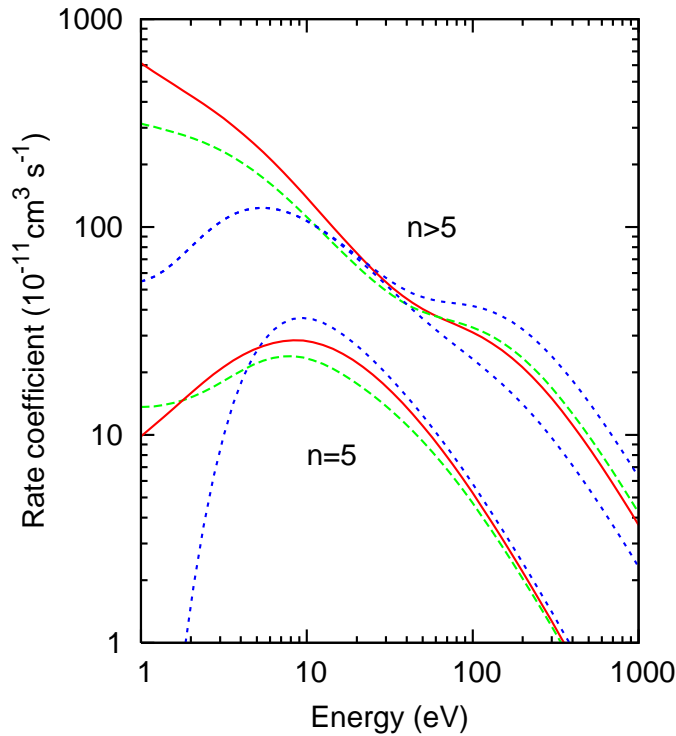}\vspace*{-4mm}
\caption{\label{Fig3}
(Color online) W$^{20+}$ Maxwellian DR rate coefficient contributions for $4f\rightarrow 5l\,(l=0-4)$ promotions 
(capture to $n=5$ and $n>5$ are shown separately):
IC (solid red curves), LS (long-dashed green curves), and CA, $n>5$ both with and without the $n=5$ continuum (short-dashed blue curves).}
\end{figure}

\begin{figure}[tbp]
\includegraphics[scale=0.8, angle=-90]{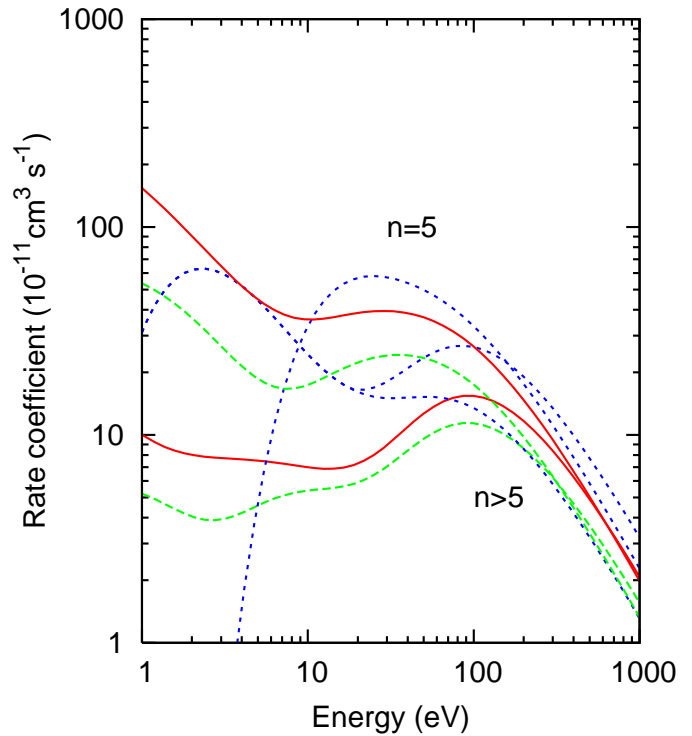}\vspace*{-4mm}
\caption{\label{Fig4}
(Color online) W$^{20+}$ Maxwellian DR rate coefficient contributions for $4d\rightarrow 5l\,(l=0-4)$ promotions 
(capture to $n=5$ and $n>5$ are shown separately):
IC (solid red curves), LS (long-dashed green curves), and CA, $n>5$ both with and without the $n=5$ continuum (short-dashed blue curves).}
\end{figure}

\subsection{$\Delta n=0$}
We present and compare our $LS$-coupling and IC results for the $4f\rightarrow 4f$ and $4d\rightarrow 4f$ promotions
in Fig.~\ref{Fig2}. We note the close agreement between them and the CA results for the $4d\rightarrow 4f$ promotion
above $\sim 100$~eV. There is about a factor 7 difference between the $LS$-coupling and IC results down at 1~eV.
The contribution from the $4f\rightarrow 4f$ is no more than about 10\% of the $4d\rightarrow 4f$ above 1~eV.

\subsection{$\Delta n=1$}

\subsubsection{$4f\rightarrow 5l$}
We present and compare our $LS$-coupling and IC results for the $4f\rightarrow 5l$ promotions
in Fig.~\ref{Fig3}. We separate-out the contributions from capture to $n=5$ and $n>5$. The sum of the two
in $LS$-coupling and IC agree to within 10\% by 100~eV. We show CA results both with and without 
autoionization to the $n=5$ continuum. The rather pronounced high energy peak
is reduced by a factor of 2 at 160~eV. These are the first autoionizations into excited states
pathways in the CA coupling scheme. The $LS$-coupling and IC are already suppressed by autoionization
into (the continuum of) a multitude of excited states within the ground configuration.
This is reflected in their less pronounced high energy peaks. We would not expect the ($n>5$) $LS$-coupling
and IC results to be suppressed further by more than $\sim$20\% below 300~eV.

\subsubsection{$4d\rightarrow 5l$}
We present and compare our $LS$-coupling and IC results for the $4d\rightarrow 5l$ promotions
in Fig.~\ref{Fig4}. We see that the relative contributions from capture to $n=5$ and $n>5$ are reversed
compared to the $4f\rightarrow 5l$ promotions. The $n>5$ contribution does not exceed that of the $n=5$ 
until high energy. This is due to autoionization suppression via the $4f\rightarrow 4d$ inner-shell transition. 
The CA results for $n>5$ are suppressed by a factor of two at 160~eV but the sum including $n=5$ by about one third.
The ($n>5$) $LS$-coupling and IC results are likely to be suppressed by a larger relative factor than
for the case of $4f\rightarrow 5l$ promotions but the overall sum including $n=5$ dilutes the factor
and $\sim 20$\% appears to be a reasonable estimate here.

\begin{figure}[tbp]
\includegraphics[scale=0.8, angle=-90]{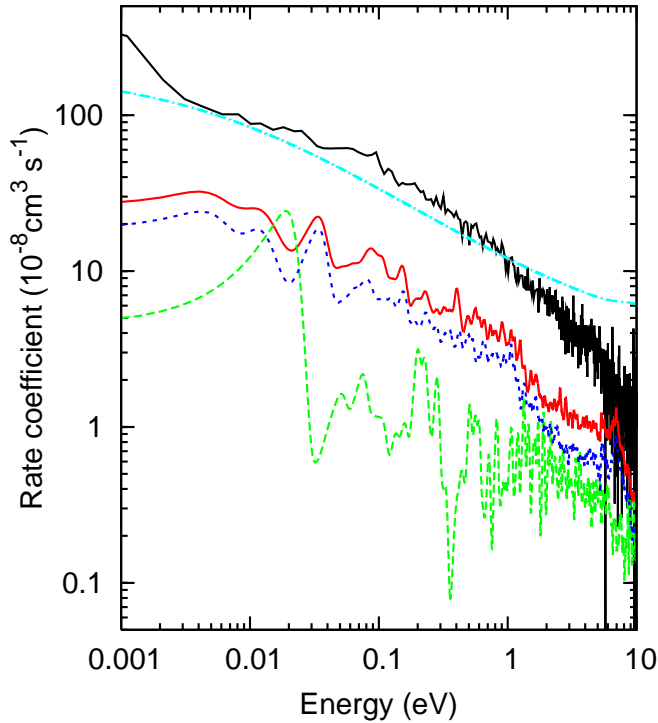}\vspace*{-4mm}
\caption{\label{Fig5}
(Color online) W$^{20+}$  merged-beams DR rate coefficients:
experiment~\cite{Sch} (solid black curve), partitioned total (dot-dashed cyan curve), IC total (solid red curve),
LS total (long-dashed green curve), and IC $4d\rightarrow 4f$ only (short-dashed blue curve).}
\end{figure}

\begin{figure}[tbp]
\includegraphics[scale=0.8, angle=-90]{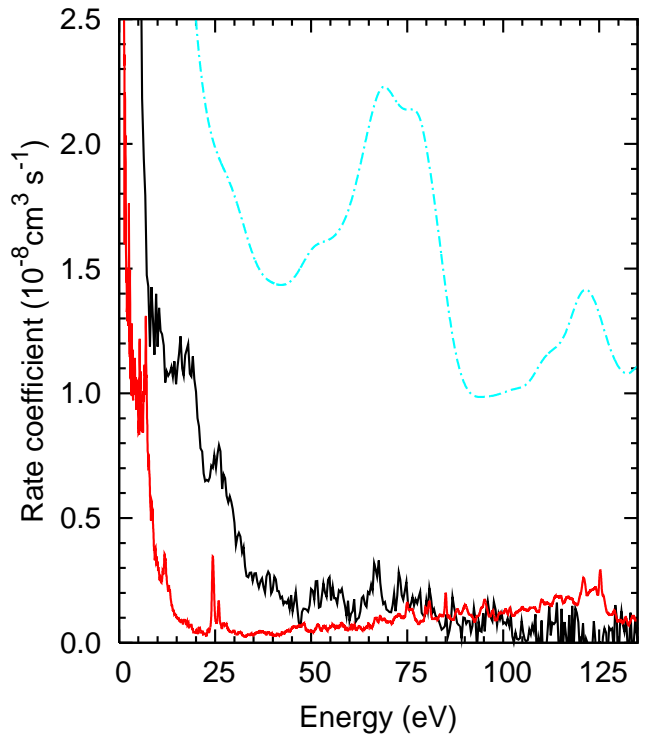}\vspace*{-4mm}
\caption{\label{Fig6}
(Color online) W$^{20+}$  merged-beams DR rate coefficients:
experiment~\cite{Sch} (solid black curve), partitioned total (dot-dashed cyan curve), IC total (solid red curve).}
\end{figure}

\begin{figure}[tbp]
\includegraphics[scale=0.8, angle=-90]{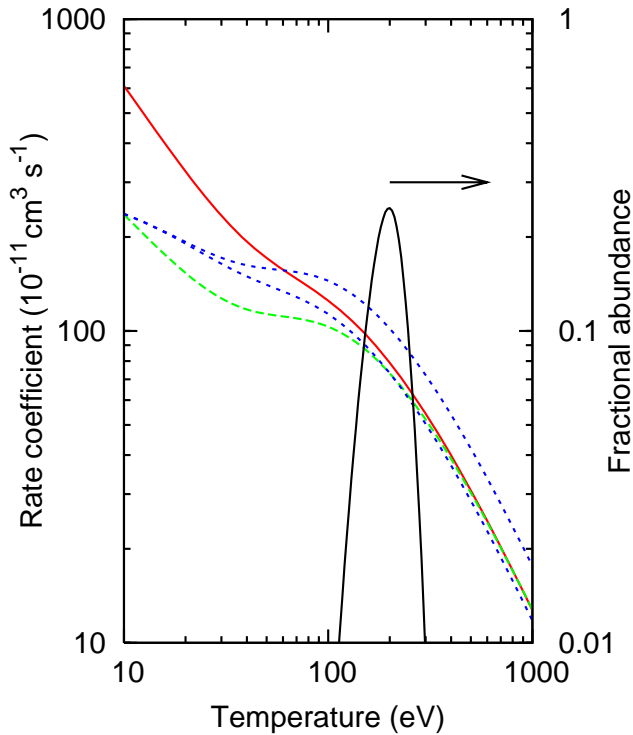}\vspace*{-4mm}
\caption{\label{Fig7}
(Color online) W$^{20+}$ total Maxwellian DR rate coefficients:
IC (solid red curve), LS (long-dashed green curve), and CA with-and-without $n=5$ continuum
(short-dashed blue curves).
The fractional abundance of W$^{20+}$ in a magnetic fusion plasma is shown also (solid black curve).}
\end{figure}

\begin{figure}[tbp]
\includegraphics[scale=0.8, angle=-90]{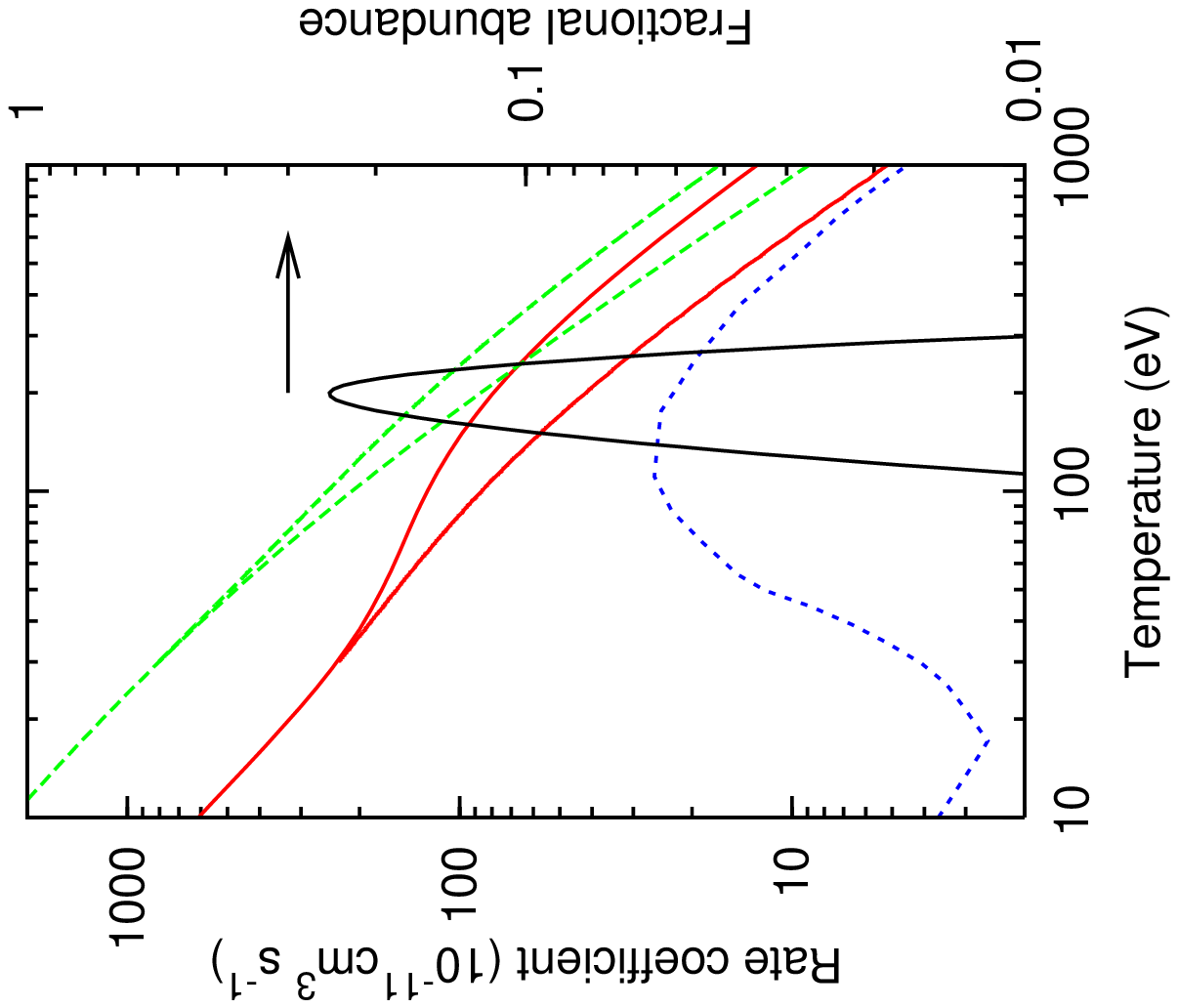}\vspace*{-4mm}
\caption{\label{Fig8}
(Color online) W$^{20+}$ total Maxwellian DR rate coefficients:
IC all resonances and to 140~eV only (solid red curves), experiment~\cite{Sch} to 140~eV and with 
theory top-up for resonances above~140eV (long-dashed green curves), and ADAS~\cite{AF} (short-dashed blue curve).
The fractional abundance of W$^{20+}$ in a magnetic fusion plasma is shown also (solid black curve).}
\end{figure}

\subsection{Totals (Merged-beams)}
We convoluted our DR cross sections with the electron velocity distribution applicable for the merged 
electron-ion beams in the TSR cooler~\cite{Sch}. We compare our resultant total DR rate coefficients
with those measured on the TSR storage ring~\cite{Sch}. We focus first on the near threshold region
[0--10]~eV for which Schippers \etal reported the largest measured DR rate coefficient to-date. 
We see in Fig.~\ref{Fig5} that our IC results are a factor 2--5 larger than our $LS$-coupling ones but they are
still typically a  factor of 3 smaller than experiment. Our IC results are dominated by the
$4d\rightarrow 4f$ promotion (80\%).  The density of resonances is such that there is little
resultant structure in the total. We have a near quasi-continuum of resonances. It does not 
matter then just where the ionization limit lies. 

The remaining factor of $\sim 3$ is likely due to the incomplete mixing within our IC configuration expansion. 
We show results (Fig.~\ref{Fig5}) where we have partitioned our autoionization rates using the Breit-Wigner
distribution with a spreading width of 10~eV~\cite{Gleb1}. The results are not particular sensitive to this width. 
Our CA, $LS$-coupling and IC results are barely distinguishable on this scale and so we show a single curve.
We obtain agreement with experiment to within 20\% over 0.003 -- 2~eV. 
Similar findings were obtained by Flambaum \etal~\cite{Flam1} for Au$^{25+}$.
(The measured cross section increases at energies below 0.003~eV due to an artifact of the merged-beams technique~\cite{Gwin}.)

The first excited level of the ground term opens-up just below 2~eV. The experimental cross section falls away progressively 
thereafter --- see Fig.~\ref{Fig6}. This fall-off coincides with an increasing number of alternative autoionization channels opening-up.
If autoionization into excited levels is fully redistributed as per the ground then the total width is unaffected. 
The partitioned autoionization widths to the ground level are orders of magnitude smaller than the radiative widths. 
This is why the partitioned results are largely unchanged --- even when summing over autoionization to hundreds of excited levels. 
In nature it would appear that typical autoionizing widths to the ground level are no more than a factor $\sim 10$ 
smaller than the radiative widths. Summing over a relatively small number of excited continua then produces a
total autoionization width comparable with and then exceeding the radiative width. 
We note that our level-resolved autoionization widths are only typically at most a factor 5--10 larger than the radiative widths. 
Experiment falls between the two theoretical `limits'.  The partitioned results are clearly inapplicable here.

All of our results are an upper limit because we assume 100\% of 
the W$^{20+}$ initial ion population to be residing in the ground state. Schippers \etal~\cite{Sch} 
identify several possible metastable levels that could remain populated during the lifetime of the 
measurement but have no estimate of their population. Any combination of metastables that we take
reduces the total. This is because DR from excited states is suppressed by autoionization into 
the continuum attached to lower levels. Only the ground level is immune from such. The agreement between
our Breit-Wigner partitioned results and experiment below 10~eV would indicate that the metastable presence
is not significant.

\subsection{Totals (Maxwellian)}
We turn now to the corresponding total Maxwellian DR rate coefficients. We compare results in Figs.~\ref{Fig7} and  \ref{Fig8}.
The two CA results (Fig.~\ref{Fig7}) illustrate the effect of omission of $n=5$ continuum suppression on the total.
It is 25\% at 160~eV.  The lower CA, $LS$-coupling, and IC results are all in close agreement
above $\sim$100~eV. The $LS$-coupling and IC results are an upper limit not just because of the omission  of the 
$n=5$ continuum but also because they assume 100\% of the W$^{20+}$ initial ion population to be residing in the ground state. 

The experiment by Schippers \etal~\cite{Sch} only detects resonances which occur below 140~eV.
We show a second IC result (Fig.~\ref{Fig8}) which imposes such a cut-off. This cut-off result lies just over 50\% below
experiment at 160~eV. If we add a theoretical top-up (for the resonances above 140~eV) to the experimental result then this
difference is reduced to about 40\% (at 160~eV) and is 50--30\% over 100--300~eV. The top-up is between 20--60\% of the 
experimental result alone over 100--300~eV. The topped-up experimental result is the total (zero-density) DR rate coefficient
that we recommend for use in modeling because of the remaining difference between theory and experiment. This difference
is attributable to the difference in the contribution from low energy resonances. The $\Delta n=0$ resonances
as a whole contribute about one third of the IC total at 160~eV. The factor 3 larger experimental
contribution from such resonances at low energies means that they contribute significantly more here.

The final results we show in Fig.~\ref{Fig8} are those from ADAS~\cite{AF}. These were
determined using the Burgess General Formula (GF)~\cite{GF} --- this is ADAS Case~A~\cite{ADAS} which extends
the GF validity to finite density by the use of a global suppression factor~\cite{GFD}. 
Those shown here were determined at close to zero electron density ($10^{8}$~cm$^{-3}$).

We see that the ADAS~\cite{AF} results lie between a factor 10 and 4.5 below our recommended ones over 100--300~eV.
The ADAS results are for dielectronic plus  radiative recombination. The radiative contribution dominates in these 
results below 20~eV because the Case~A Burgess GF cannot describe the effect of low-lying resonances. Such low-lying
resonances can be described by the the BBGP approach~\cite{Burg,BBGP} --- this is ADAS Case~B which resolves the final
recombined state and so is amenable to the collisional-radiative modeling of density effects~\cite{CR}.

The fractional abundance curve we show has been determined using the ADAS Case~A data at an electron density
of $10^{13}$~cm$^{-3}$ which is typical of that relevant to magnetic fusion edge plasmas~\cite{Putt}. It differs slightly
from the one shown by Schippers \etal~\cite{Sch} which is due to P\"{u}tterich~\cite{Putt2} and at $10^{14}$~cm$^{-3}$. 
It is appropriate to use a finite-density abundance to indicate the plasma relevant temperatures on which
to focus our comparisons of zero-density rate coefficients because the temperature of peak abundance is
density sensitive.
Use of rate coefficients which are up to a factor 10 larger though is likely to move the peak abundance to 
higher temperature.
(Similar increases in the DR rates to used can be expected for adjacent ionization stages.)

\subsubsection{Density effects}
A rigorous treatment of density effects on dielectronic recombination rate coefficients and their
consequential effect on the ionization balance of W$^{20+}$ and adjacent ionization stages is
beyond the scope of the present work. We can make some observations though.
The ADAS rate coefficient is reduced by a factor of 2 at an electron density
of $10^{13}$~cm$^{-3}$ (not shown) compared to zero-density. This is due to the stepwise ionization of
high-$n$ ($\gtrsim 10$) Rydberg states following recombination.  The new recommended total DR rate coefficients
contain a large contribution from low-energy resonances of low-$n$ ($\lesssim 10$). If we assume that the high-$n$ 
contributions to both are broadly similar then we can expect maybe a 10--20\% reduction in the new recommended values over 100--300~eV.
Similar (reduced) effects can be expected for adjacent ionization stages. This means that the corresponding fractional
abundances are likewise less sensitive to the electron density than indicated by the current ADAS data.
A revision of the density dependent ionization balance of f-shell tungsten ions is clearly needed.

\subsubsection{Fitting coefficients}
It is convenient often for simple modeling purposes to fit the total Maxwellian 
dielectronic recombination rate coefficient ($\alpha$) to the following functional form:
\begin{eqnarray}
\alpha(T)&=&T^{-3/2}\sum_i c_i \exp\left(\frac{-E_i}{T}\right) \nonumber
\label{DRfit}
\end{eqnarray}
where the $E_i$ are in the units of temperature $T$ (e.g. eV) and the units of $c_i$ are 
then cm$^3$s$^{-1}$[eV]$^{3/2}$.

\begin{table}
\caption
{Recommended total (zero-density) dielectronic recombination rate coefficient fitting coefficients $c_i$~(cm$^3$s$^{-1}$[eV]$^{3/2}$)
and $E_i$(eV) for the initial ground  level of W$^{20+}$.\label{tab}}
\begin{ruledtabular}
\begin{tabular}{llll}
$  i$ & $c_i $ & $ E_i$\\
\hline
  1 & 4.025($-$7)\footnote{(m) denotes $\times 10^m$.}&    1.093(+0)\\
  2 & 7.697($-$7)&    9.153(+0)\\
  3 & 1.065($-$6)&    3.425(+1)\\
  4 & 1.487($-$6)&    1.205(+2)\\
  5 & 2.177($-$6)&    2.384(+2)\\
\end{tabular}
\end{ruledtabular}
\end{table}

In Table \ref{tab} we present such fitting coefficients for the recommended (experiment topped-up by theory)
total zero-density dielectronic recombination rate coefficient for the initial ground 
level of W$^{20+}$. The fit is accurate to better (often much better) than 1\% over 1--1000~eV.

The total dielectronic recombination rate coefficient can be taken to be the
total recombination rate coefficient. The contribution from radiative
recombination is negligible over the given temperature range as is that from three-body
recombination at the densities of interest to magnetic fusion plasmas.

\section{Conclusion}
\label{Con}

We have calculated IC DR rate coefficients for W$^{20+}$ which include all significant one-electron promotions-plus-capture.
A factor 3 difference with experiment remains at low energies. We have demonstrated that this can be
removed if we assume complete chaotic mixing of multiply-excited near-threshold configurations.
A similar finding was obtained by Flambaum \etal~\cite{Flam1} for Au$^{25+}$.
The difference between theory and (topped-up) experiment at energies relevant to magnetic fusion modeling for ITER is
somewhat less viz. between a factor two and 1.5 over 100--300~eV. The DR data used by ADAS for such modeling 
needs to be updated since the current Burgess GF Case~A results lie between a factor 10 and 4.5 below our 
new recommended values over 100--300~eV.
Our CA, $LS$-coupling and IC results are all in close accord (20\%) above 100eV which suggests that DR rate
coefficients for complex W ions can be determined readily to within a factor of two for modeling purposes.
Similar behavior can be expected for related complex ions of other heavy elements. Determination of such DR rate
coefficients which are accurate to the $\sim$~20\% level remains problematic though.

\section*{ACKNOWLEDGMENTS}
We would like to thank Stefan Schippers for providing the experimental data in numerical form.
One of us (NRB) would like to thank Gleb Gribakin for clarifying the methodology used in~\cite{Flam1}.
This work was supported in part by a Euratom Framework 7 Support Action Agreement with the
University of Strathclyde and US DoE grants to Auburn University.


\end{document}